\documentclass[aps,prx,reprint,showpacs,floatfix,superscriptaddress]{revtex4-2}
\usepackage{graphicx}
\usepackage{amsmath,amsfonts,amscd, physics}
\usepackage{inputenc}
\usepackage{mathrsfs, dsfont}
\usepackage{mathtools}
\usepackage{stmaryrd}
\SetSymbolFont{stmry}{bold}{U}{stmry}{m}{n}
\usepackage{array}
\usepackage{bm}
\usepackage{setspace}
\usepackage[T1]{fontenc}
\usepackage{txfonts}
\usepackage{lineno}
\usepackage{csquotes}
\usepackage{xcolor}
\usepackage{cuted}
\usepackage{flushend}

\begin{document}

\title{Exact Quantum Fisher Matrix Results for Distributed Phases Using Multiphoton Polarization Greenberger–Horne–Zeilinger States}
\author{Jiaxuan Wang}
\email{jxnwang@tamu.edu}
\affiliation{Department of Physics and Astronomy, Texas A\&M University, College Station, Texas 77843, USA}
\affiliation{Institute for Quantum Science and Engineering, Texas A\&M University, College 
Station, Texas 77843, USA}
\author{Girish S. Agarwal}
\affiliation{Department of Physics and Astronomy, Texas A\&M University, College Station, Texas 77843, USA}
\affiliation{Institute for Quantum Science and Engineering, Texas A\&M University, College Station, Texas 77843, USA}
\affiliation{Department of Biological and Agricultural Engineering, Texas A\&M University, College Station, Texas 77843, USA}

\begin{abstract}

In recent times, distributed sensing has been extensively studied using squeezed states. 
While this is an excellent development, it is desirable to investigate the use of other quantum probes, such as entangled states of light. 
In this study, we focus on distributed sensing, i.e., estimating multiple unknown phases at different spatial nodes using multiphoton polarization-entangled Greenberger–Horne–Zeilinger (GHZ) states distributed across different nodes.
We utilize tools of quantum metrology and calculate the quantum Fisher information matrix (QFIM). 
However, the QFIM turns out to be singular, hindering the determination of quantum Cramér-Rao bounds for the parameters of interest. 
Recent experiments have contended with a weaker form of the Cramér-Rao bound, which does not require the inversion of the QFIM. 
It is desirable to understand how relevant these weaker bounds are and how closely they approach the exact Cramér-Rao bounds.
We thus analyze the reason for this singularity and, by removing a redundant phase, obtain a nonsingular QFIM, allowing us to derive exact quantum Cramér-Rao bounds. 
Using the nonsingular QFIM, we show that the arithmetic average of the distributed phases is Heisenberg-limited. 
We demonstrate that the quantum metrological bounds can be saturated by projective measurements, enabling us to determine the Fisher information matrix (FIM), which is also singular. We then show how this singularity can be resolved.

\end{abstract}

\maketitle

\section{Introduction}

In recent decades, quantum metrology has experienced substantial progress 
\cite{giovannetti2004quantum, giovannetti2011advances, polino2020photonic}, especially in distributed quantum sensing \cite{zhang2021distributed}. 
This advanced technique enables the simultaneous examination of systems across multiple spatial locations, which has become a fundamental aspect of quantum networks. 
These networks consist of interconnected quantum devices that facilitate enhanced distributed computing and secure communication \cite{appas2021flexible}.
A significant challenge in this field is achieving precise measurements for multiple parameters due to the Heisenberg uncertainty principle, which imposes fundamental limits on the accuracy of certain pairs of measurements \cite{szczykulska2016multi, busch2007heisenberg, lu2021incorporating}. 
Entanglement between quantum states at remote nodes is critical for surpassing the measurement capabilities of systems without entanglement \cite{zhuang2018distributed, 
guo2020distributed, xia2020demonstration,malia2022distributed,zhang2024entanglement}. 
The optimal probe states and measurements for multiple phase estimation are given in \cite{humphreys2013quantum}, which requires special quantum correlations to obtain the minimum total variance.
Distributing a single-mode squeezed vacuum state through a balanced beam splitter network allows for precise quadrature displacement estimation, achieving Heisenberg scaling in the number of modes \cite{zhuang2018distributed, guo2020distributed}. 
This state is proven to be the optimal entangled Gaussian state for estimating both displacement and the weighted sum of phase \cite{oh2020optimal}.
The scalability of this approach has been illustrated in global random continuous-variable (CV) networks. Kwon et al. showed that most CV quantum networks can achieve Heisenberg scaling for distributed quantum displacement sensing \cite{kwon2022quantum}. 
Additionally, these networks can maintain robustness in the presence of noise by using error correction codes, which help restore Heisenberg scaling even in networks with up to 100 nodes \cite{zhuang2020distributed}. 

Complementing these advancements in CV systems, discrete variable (DV) quantum systems have also shown significant promise in quantum sensing. 
In 2001, Pan et al. \cite{pan2001experimental} achieved a breakthrough by creating a highly pure four-photon Greenberger–Horne–Zeilinger (GHZ) state \cite{greenberger1990bell}. 
The quality of entangled DV states has further increased in recent years \cite{haffner2005scalable, leibfried2005creation,monz2011qubit, wang2016experimental,huang2011experimental,  
chen2017observation, song2017qubit, wang2018qubit, albarelli2020perspective, zhong2021deterministic,friis2018observation}, showcasing the strong potential of DV systems in quantum sensing. 
Given their notable scalability and applicability to various quantum technologies, studying the sensing precision of DV systems is essential.
Consider a system with $d$ unknown phases $\phi = {\phi_1, \phi_2, \ldots, \phi_d}$ in distinct spatial nodes, where the goal is to estimate a linear combination of these phases. 
Theoretical demonstrations \cite{gessner2018sensitivity} have shown that Heisenberg scaling is achievable using both mode-entangled and particle-entangled states.
This has been experimentally validated for mode numbers $d=3$ \cite{liu2021distributed} and $d=2$ up to 10 km \cite{zhao2021field}. 
However, to estimate multiple unknown phases with those states, the number of photons $N$ must be equal to or greater than $d$. 
Kim et al. \cite{kim2024distributed} introduced the polarization GHZ state for multi-phase estimation, achieving Heisenberg scaling without requiring $N \geq d$. 
This approach enhanced sensitivity by 2.2 dB for $N=2$ and $d=4$ over 3 km.

Understanding parameter estimation precision in these quantum systems relies on the quantum Fisher information (QFI) and quantum Cramér-Rao bound (QCRB) \cite{fisher1922dominance, cramer1946methods, braunstein1992quantum, braunstein1994statistical, fujiwara1994information, helstrom1967minimum, helstrom1968minimum, helstrom1969quantum}, which are essential for discerning quantum sensing capabilities \cite{toth2014quantum}. The quantum Fisher information matrix (QFIM) \cite{fujiwara1995quantum, petz1996geometries, petz1996monotone, ercolessi2012geometry, ercolessi2013symmetric, contreras2016on, liu2020quantum, albarelli2020perspective} provides insight into parameter correlations and achievable precision. 
In single-parameter estimation, the QCRB is attainable, and there are well-established methods for identifying the optimal measurement \cite{braunstein1994statistical, braunstein1996generalized}.
However, in the multi-parameter scenario, the QCRB for all parameters can not be attained simultaneously unless the optimal measurements for all parameters are compatible \cite{szczykulska2016multi}. 
Special conditions are required to determine whether the QCRB can be achieved by certain measurements \cite{matsumoto2002new, humphreys2013quantum, pezze2017optimal}.
Trade-offs arise when dealing with incompatible quantum measurements \cite{ozawa2004uncertainty, branciard2013error, roccia2018multiparameter, xia2023toward}.
In cases with limited parameter dependency understanding, machine learning techniques like Bayesian quantum estimation are employed \cite{nolan2021machine,fiderer2021neural,cimini2023deep}.
Probabilistic protocols with post-selection are proposed to achieve the Heisenberg limit for general network states \cite{yang2024quantum}.
In previous studies, such as \cite{kim2024distributed}, the analysis of multi-partite GHZ states focused only on the weaker form of the Cramér-Rao bound (CRB), even though complete system information was available. 
A similar limitation was observed in \cite{liu2021distributed}, where the systems with mode-entangled states only allowed for the estimation of a combination of phases, making it impossible to determine individual phases. 
In our study, we explore the full quantum metrological framework and demonstrate that the difficulty in obtaining the exact CRB stems from the singularity of the Fisher information matrix (FIM). This singularity arises because the phases \(\{\phi_1, \phi_2, \ldots, \phi_d\}\) do not form an independent basis when using polarization GHZ states for estimation, necessitating a proper variable transformation.

We detail the process for constructing these variable transformations, which result in diagonal or block-diagonal quantum Fisher information matrices. With this approach, we derive the quantum Cramér-Rao bounds for the problem and calculate the achievable classical CRB based on probability measurements. Notably, the QCRB for the average phase \(\bar{\phi} = \sum_{i=1}^d \phi_i / d\) is \(1/N\), which achieves Heisenberg scaling for any even number \(d\).
This method enables us to achieve optimal precision, providing new insights into quantum techniques for enhanced sensing in distributed measurements.
This paper is structured as follows: 
In Sec.~II, we begin with an illustrative example by focusing on the case where $N=2$ and $d=4$, considering the average phase $\bar{\phi} = \frac{1}{d}(\phi_1 + \phi_2 + ... + \phi_d)$ as the parameter to be estimated. 
Sec.~III discusses the singularity of the QFIM and the parameter transformation needed to derive the QCRB. 
In Sec.~IV, we examine the projective measurement scheme and obtain the classical Fisher information matrix. 
Sec.~V follows the given parameter transformation and obtains the exact Cramér-Rao bound for the estimation of $\bar{\phi}$.
Sec.~VI extends the analysis to generalize the relationship between precision and the number of photons and phases, confirming Heisenberg scaling. 
Sec.~VII considers projective measurements on this general system, demonstrating that the sensitivity for the average phase achieves the QCRB.
In Sec.~VIII, we discuss the relation between the weak Cramér-Rao bound and the exact Cramér-Rao bound.
Finally, Sec.~IX offers a discussion of our findings and their implications for quantum sensing.

\section{Distribution of two-photon entanglement across four nodes and Quantum Fisher Information}

As an example, we consider the scenario where the input photon number $N=2$ and the number of spatially separated phases $d=4$.
A polarization GHZ state of two photons, generated via spontaneous parametric down-conversion (SPDC), can be expressed by
\begin{equation} \label{phiab}
\ket{\Phi_{a,b}}=\frac{1}{\sqrt{2}}(\ket{H_aH_b}+\ket{V_aV_b}),
\end{equation}
where $a$ and $b$ denote each photon in the two-photon GHZ state; 
$H$ and $V$ denote the polarization states of the photons, representing horizontal and vertical orientations, respectively. 
By adding a beam splitter network that contains two 50:50 beam splitters, the two-photon GHZ state can be split to distribute among four nodes, thus forming the input state of this sensing problem, expressed by
\begin{equation} \label{phi_i}
\ket{\Phi_i}=\frac{1}{2}(\ket{\Phi_{1,2}}+\ket{\Phi_{2,3}}+\ket{\Phi_{3,4}}+\ket{\Phi_{4,1}}).
\end{equation}
Here, the different subscripts refer to different spatial nodes. After passing through various phases at different nodes, the input state is transferred to the following output state
\begin{align} \label{phi_o}
\ket{\Phi_o}=&\frac{1}{2\sqrt{2}}(\ket{H_1H_2}+e^{i(\phi_1+\phi_2)}\ket{V_1V_2}\nonumber\\
&+\ket{H_2H_3}+e^{i(\phi_2+\phi_3)}\ket{V_2V_3} \nonumber\\
&+\ket{H_3H_4}+e^{i(\phi_3+\phi_4)}\ket{V_3V_4}\nonumber\\
&+\ket{H_4H_1}+e^{i(\phi_4+\phi_1)}\ket{V_4V_1}).
\end{align}

When analyzing such systems, precision becomes crucial.
Precision in parameter estimations is quantified by the standard deviation ($\Delta X$) of measured parameter values: $\Delta X=\sqrt{\langle X^2\rangle-\langle X\rangle^2}$. 
In systems with multiple parameters, the precision of determining each parameter's value can be influenced not only by their individual characteristics but also by the covariance between different parameters, which may not be independent.
To establish a classical lower bound for this precision, we utilize the classical Fisher information matrix $F_{mn}(X)$. This matrix quantifies the information content about the parameters $X_m$ and $X_n$ present in the data. Mathematically, it is defined as the expectation value of the second derivative of the log-likelihood function with respect to these parameters:
\begin{equation} \label{fi_1}
F_{mn}(X) = \langle \frac{\partial^2}{\partial X_m \partial X_n} \ln L(X) \rangle,
\end{equation}
where $L(X)$ represents the likelihood function. For discrete data, the likelihood function is the product of the probability mass function (PMF) evaluated at each observed value:
$L(X) = \prod_j P_j(X)$,
where $P_j(X)$ is the PMF evaluated at the $j$th individual observed value given the parameter values $X$. Thus, it can be written as
\begin{equation} \label{fi_2}
    F_{mn}(X)=\sum_j\frac{1}{P_j(X)}\frac{dP_j(X)}{dX_m}\frac{dP_j(X)}{dX_n}.
\end{equation}
The lower bounds (Cramér-Rao bounds) for the variances and covariances are defined as 
\begin{equation} \label{crb_1}
    Cov(X_m,X_n)\ge F^{-1}_{mn}(X)/{\cal N},
\end{equation}
where $F^{-1}_{mn}(X)$ is the Fisher information matrix's inverse matrix, and ${\cal N}$ is the number of independent measurements.

\begin{figure}[htbp]
\centering\includegraphics[width=\columnwidth]{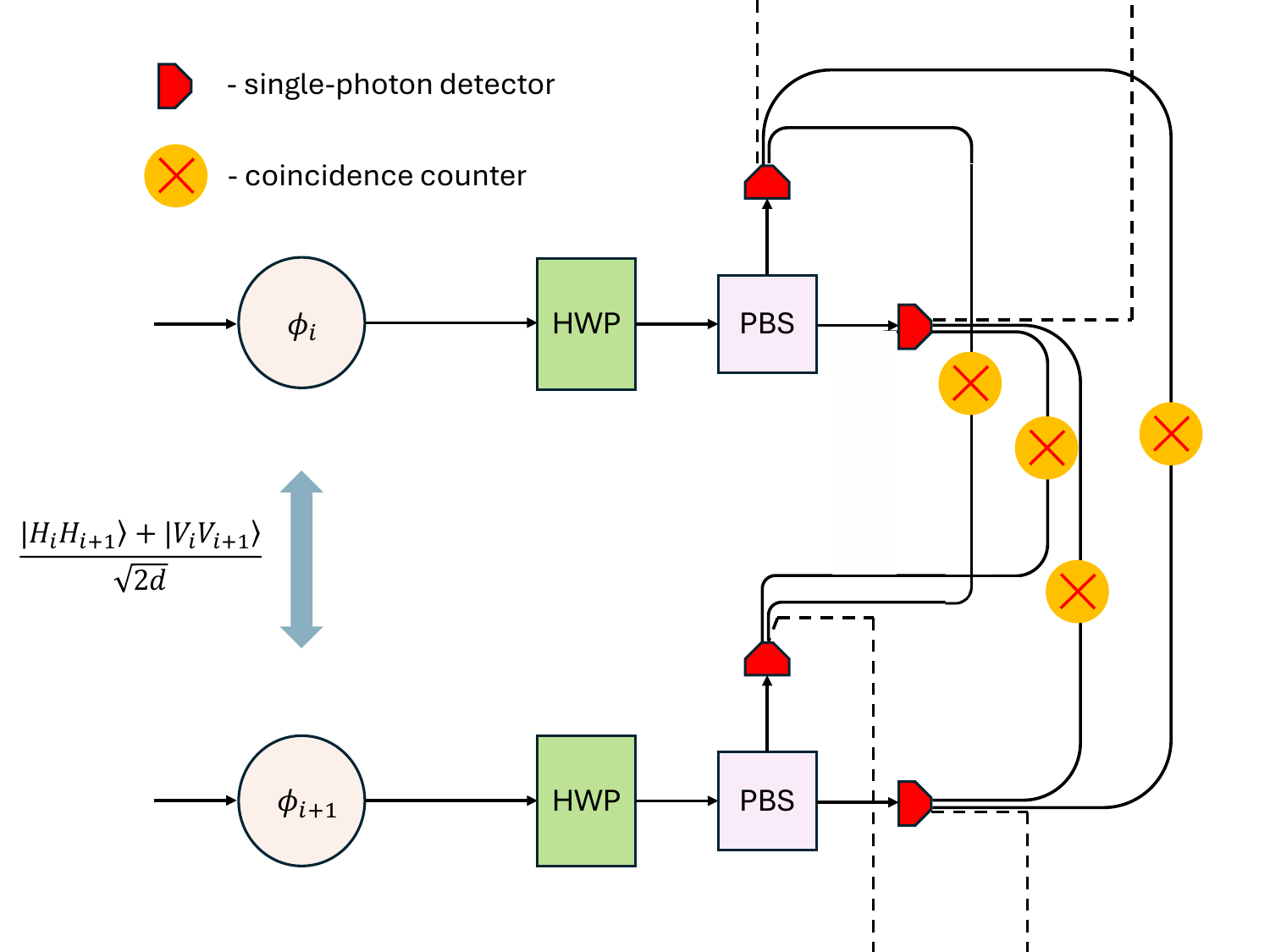}
\caption{The illustration of the projection measurement method. In this scheme, HWP denotes the half-wave plate, and PBS represents the polarization beam-splitter.} \label{projection}
\end{figure}

Expanding this framework to quantum mechanics involves maximizing $F(X)$ over all conceivable quantum measurements, leading to the quantum Fisher information (QFI) ${\cal F}^{(Q)}(X)$. Various techniques exist for obtaining the QFI. Here, given that the final state $\ket{\Phi_o}$ is a pure state, the elements in the QFIM can be calculated by
\begin{equation} \label{fq_def}
    {\cal F}^{(Q)}_{mn}(X) = 4\text{Re}\left(\langle\partial_{X_m}\psi \,|\, \partial_{X_n}\psi \rangle- \langle\partial_{X_m}\psi \,|\, \psi \rangle\, \langle\psi \,|\, \partial_{X_n}\psi \rangle\right).
\end{equation}
Thus, the corresponding quantum Cramér-Rao bounds are expressed as 
\begin{equation} \label{qcrb_1}
    Cov(X_m,X_n)\ge ({\cal F}^{(Q)})^{-1}_{mn}(X)/{\cal N}.
\end{equation}
Using Eq.~\eqref{fq_def}, we obtain the QFIM of the original set of variables as
\begin{equation}
    {\cal F}_{0}^{(Q)}=\left(\begin{array}{cccc}
\frac{3}{4} & \frac{1}{4} & -\frac{1}{4} & \frac{1}{4}\\
\frac{1}{4} & \frac{3}{4} & \frac{1}{4} & -\frac{1}{4}\\
-\frac{1}{4} & \frac{1}{4} & \frac{3}{4} & \frac{1}{4}\\
\frac{1}{4} & -\frac{1}{4} & \frac{1}{4} & \frac{3}{4}
\end{array}\right),
\end{equation}
The rank of \({\cal F}_{0}^{(Q)}\) is 3, indicating that it is a singular matrix. Consequently, it is not possible to obtain the QCRBs for the four variables, as the inverse operation required by Eq.~\eqref{qcrb_1} cannot be performed on a singular QFIM.

\section{Renormalization of the singularity of the Quantum Fisher Information Matrix and the Heisenberg Limited Sensitivity of Distributed Sensing}

To deal with the singularity, it is essential to investigate the degree of freedom in this system and identify the independent variables that contribute.
we construct an orthogonal transformation as follows: the original variables $\{\phi_1,\phi_2,\phi_3,\phi_4\}$ can be orthogonally transformed to new basis $\{\phi_0,\phi_a,\phi_b,\phi_c\}$, where
\begin{align} \label{transform_1}
\phi_{0}=\frac{\phi_{1}-\phi_{2}+\phi_{3}-\phi_{4}}{2}, \,\,\,\,
\phi_{a}=\frac{\phi_{1}+\phi_{2}+\phi_{3}+\phi_{4}}{2}, \nonumber\\
\phi_{b}=\frac{\phi_{1}+\phi_{2}-\phi_{3}-\phi_{4}}{2}, \,\,\,\,
\phi_{c}=\frac{\phi_{1}-\phi_{2}-\phi_{3}+\phi_{4}}{2}.
\end{align}
The output state can thus be expressed as
\begin{align} \label{phi_o2}
\ket{\Phi_o}=&\frac{1}{2\sqrt{2}}(\ket{H_1H_2}+e^{i(\phi_a+\phi_b)}\ket{V_1V_2}\nonumber\\
&+\ket{H_2H_3}+e^{i(\phi_a-\phi_c)}\ket{V_2V_3} \nonumber\\
&+\ket{H_3H_4}+e^{i(\phi_a-\phi_b)}\ket{V_3V_4}\nonumber\\
&+\ket{H_4H_1}+e^{i(\phi_a+\phi_c)}\ket{V_4V_1}).
\end{align}
Note that $\ket{\Phi_o}$ depends only on three variables, $\phi_a$, $\phi_b$, and $\phi_c$, while $\phi_0$ does not contribute. 

Using Eq.~\eqref{fq_def}, the QFIM of the set of variables $\{\phi_a,\,\phi_b,\,\phi_c\}$ is obtained as
\begin{equation} \label{fq_1}
    {\cal F}^{(Q)}=\left(\begin{array}{ccc}
{\cal F}^{(Q)}_{aa} & {\cal F}^{(Q)}_{ab} & {\cal F}^{(Q)}_{ac}\\
{\cal F}^{(Q)}_{ba} & {\cal F}^{(Q)}_{bb} & {\cal F}^{(Q)}_{bc}\\
{\cal F}^{(Q)}_{ca} & {\cal F}^{(Q)}_{cb} & {\cal F}^{(Q)}_{cc}
\end{array}\right),
\end{equation}
where
\begin{align*}
    {\cal F}_{aa}^{(Q)}&=\frac{7}{4}-\frac{1}{8}(\cos2\phi_{b}+\cos2\phi_{c}+4\cos\phi_{b}\cos\phi_{c}), \\
    {\cal F}_{ab}^{(Q)}&=({\cal F}_{ba}^{(Q)})^{*}=-\frac{i}{4}\cos\phi_{b}\sin\phi_{c}-\frac{i}{8}\sin2\phi_{b},\\
    {\cal F}_{ac}^{(Q)}&=({\cal F}_{ca}^{(Q)})^{*}=-\frac{i}{4}\sin\phi_{b}\cos\phi_{c}-\frac{i}{8}\sin2\phi_{c},\\
    {\cal F}_{bb}^{(Q)}&=\frac{7}{8}+\frac{1}{8}\cos2\phi_{b},\\
    {\cal F}_{cc}^{(Q)}&=\frac{7}{8}+\frac{1}{8}\cos2\phi_{c},\\
    {\cal F}_{bc}^{(Q)}&={\cal F}_{cb}^{(Q)}=-\frac{1}{4}\sin\phi_{b}\sin\phi_{c}.
\end{align*}
Considering weak phase shifts that $\phi_{a,b,c}\ll 1$, ${\cal F}^{(Q)}$ becomes the
3-dimensional identity matrix $I_3$. Thus, the quantum Cramér-Rao bound for $\bar{\phi}$ is
\begin{equation} \label{qcrb_ave}
\Delta_q \bar{\phi}=\frac{1}{2} \Delta_q \phi_a \geq\frac{1}{2},
\end{equation}
which achieves the Heisenberg scaling $\Delta \bar{\phi}\geq 1/N$ with a two-photon input.
Note that Eq.\eqref{fq_1} remains valid for any combination where $\phi_b=\phi_c=0$, i.e., as long as $\phi_1 = \phi_3$ and $\phi_2 = \phi_4$, the Quantum Cramer-Rao Bounds (QCRBs) for $\phi_a, \phi_b,$ and $\phi_c$ achieve the Heisenberg scaling. Additionally, for cases where $\phi_c = \pm \phi_b = n\pi$ for any integer $n$, Eq.\eqref{fq_1} still holds. 

\section{Measurement of Sensitivity: Classical Fisher Information Matrix}

Considering a projective measurement on the $\sigma_x$ basis shown in Fig~\ref{projection},
the corresponding probability set comprises sixteen probabilities denoted as ${P_{jk}^{++},P_{jk}^{--},P_{jk}^{+-},P_{jk}^{-+}}$. Here, the superscripts $+$ and $-$ signify the two different basis $\frac{1}{\sqrt{2}}(\ket{H}\pm\ket{V})$, and $jk$ indicates the nodes based on $\phi_{j}$ and $\phi_{k}$, i.e., ${jk} = \{12, 23, 34, 41\}$. For instance, $P_{12}^{+-}$ represents the probability of two-photon coincidence events between observing $+$ in node 1 and $-$ in node 2. 
The probabilities for the projection measurement are indicated as
\begin{align} \label{pjk_1}
P_{jk}^{++}=P_{jk}^{--}=\frac{1+\cos(\phi_{j}+\phi_{k})}{16},\nonumber \\
P_{jk}^{+-}=P_{jk}^{-+}=\frac{1-\cos(\phi_{j}+\phi_{k})}{16}.
\end{align}
Kim et al. \cite{kim2024distributed} obtained the classical Fisher information matrix for this system as
\begin{equation} \label{fim_0}
    F_0=\left(\begin{array}{cccc}
F_{11} & F_{12} & F_{13} & F_{14}\\
F_{21} & F_{22} & F_{23} & F_{24}\\
F_{31} & F_{32} & F_{33} & F_{34}\\
F_{41} & F_{42} & F_{43} & F_{44}
\end{array}\right)=\left(\begin{array}{cccc}
\frac{1}{2} & \frac{1}{4} & 0 & \frac{1}{4}\\
\frac{1}{4} & \frac{1}{2} & \frac{1}{4} & 0\\
0 & \frac{1}{4} & \frac{1}{2} & \frac{1}{4} \\
\frac{1}{4} & 0 & \frac{1}{4} & \frac{1}{2}
\end{array}\right).
\end{equation}
Note that the rank of $F_0$ is 3. Thus, an inverse operation can not be applied on $F_0$ due to the singularity. 
They then investigated the precision bound for measuring the average phase  $\bar{\phi}=\frac{1}{4}(\phi_1+\phi_2+\phi_3+\phi_4)$ using the weak form of the Cramér-Rao bound, which for a linear combination of the input $\alpha^T\phi=\sum_{i=1}^d\alpha_i\phi_i$ where $\sum_{i=1}^d|\alpha_i|=1$, is defined as
\begin{equation}\label{weakCRB}
    Var(\alpha^T\phi)\ge \frac{(\alpha^T\alpha)^2}{{\cal N}\alpha^TF_0\alpha}.
\end{equation}
\noindent Unlike Eq.~\eqref{crb_1}, this approach does not necessitate the inversion of the FIM.

\section{Removal of Singularity of the classical Fisher Information Matrix and the Cramér-Rao bound}

Considering the same measurement scheme as in Eq.~\eqref{pjk_1}, yet after constructing the transformation in Eq.~\eqref{transform_1}, we can write Eq.~\eqref{pjk_1} with the new set of variables.
Considering weak phase shifts that $\phi_{a,b,c}\ll 1$, the classical Fisher information matrix is obtained as
\begin{equation} \label{fim_1}
    F=\left(\begin{array}{ccc}
F_{aa} & F_{ab} & F_{ac}\\
F_{ba} & F_{bb} & F_{bc}\\
F_{ca} & F_{cb} & F_{cc}
\end{array}\right)=\left(\begin{array}{ccc}
1 & 0 & 0\\
0 & \frac{1}{2} & 0\\
0 & 0 & \frac{1}{2}
\end{array}\right),
\end{equation}
which is a $3\times 3$ full-rank diagonal matrix. Thus, the classical Cramér-Rao bound $\Delta_c \bar{\phi}$ for $\bar{\phi}=\frac{1}{2}\phi_a$, obtained by using the nonsingular Fisher information matrix, is
\begin{equation} \label{ccrb_1}
\Delta_c \bar{\phi}=\frac{1}{2} \Delta_c \phi_a \geq\frac{1}{2}.
\end{equation}
Furthermore, it demonstrates that the weak bound obtained in reference \cite{kim2024distributed}  attains the exact Cramér-Rao bound for the average phase in this scenario. 
It's noteworthy that the result specifically pertains to the average $\bar{\phi}$ here, where both the classical and quantum bounds coincide. This suggests that the projection measurement in \cite{kim2024distributed} serves as an optimal measurement scheme for estimating $\bar{\phi}$.
However, the quantum Cramér-Rao bound is below the counterpart when our interest extends beyond $\bar{\phi}$ to encompass parameters such as $\phi_b$, $\phi_c$, or other linear combinations within ${\phi_1,\phi_2,\phi_3,\phi_4}$, as shown in Eq.~\eqref{fim_1}. Other measurement schemes should be considered in that case.

\section{N-photon entangled states on a network with \MakeLowercase{d} nodes}

\subsection{Symmetry of the system}

\begin{figure}[htbp]
\centering\includegraphics[width=\columnwidth]{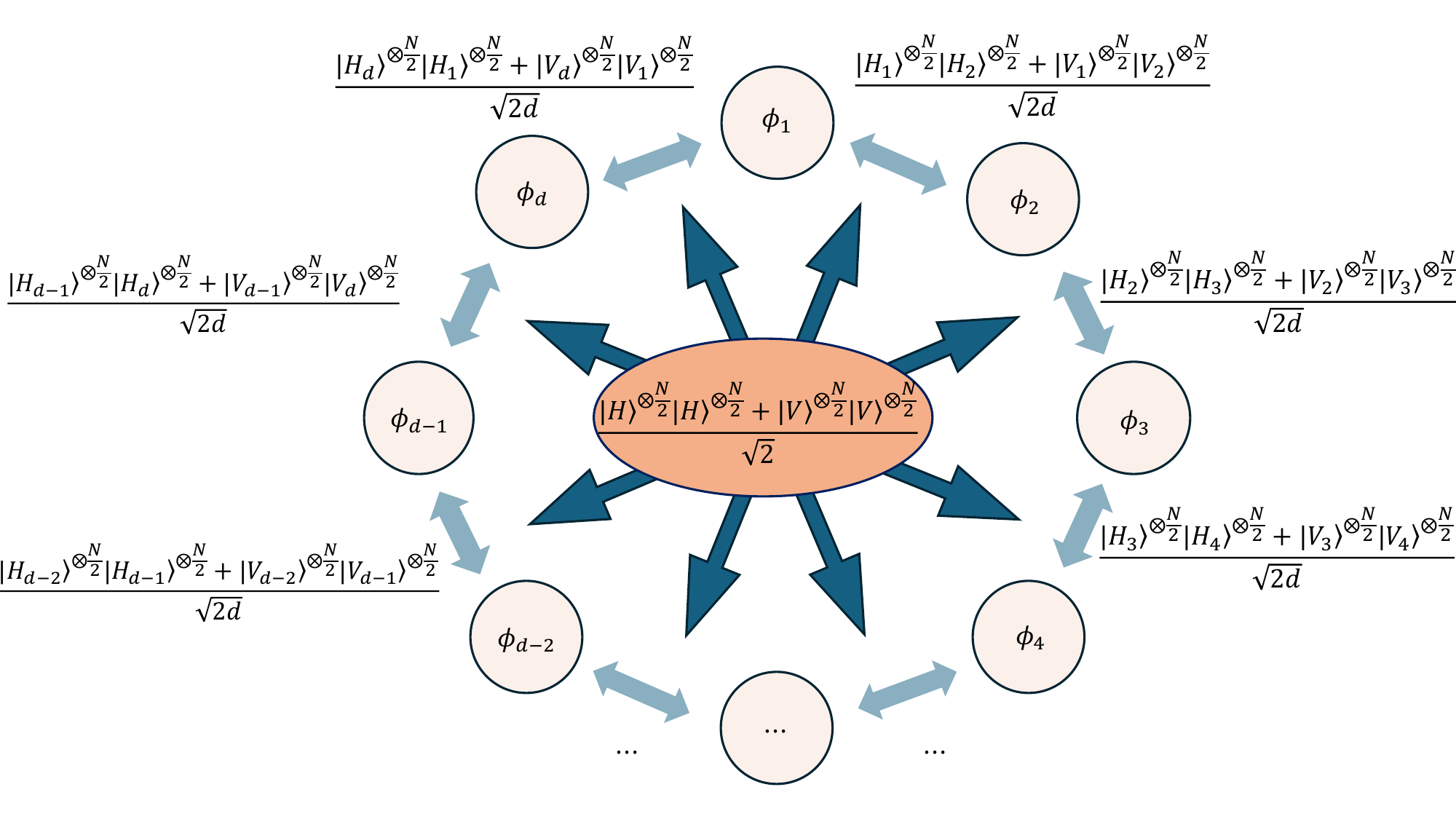}
\caption{State distribution for measurement of phases located in distinct spatial nodes.} \label{distribute}
\end{figure}

To further study the relationship between precision and photon number in this system,
we consider the most general case containing d independent phases $\{\phi_1,\ldots,\phi_d\}$ located at various spatial positions. 
Using a beam splitter network equipped with an increased number of beam splitters allows for the investigation of such distributed systems with a greater number of spatial nodes. 
Upon distribution, as shown in Fig.~\ref{distribute}, the N-photon polarization Bell state is expressed by
\begin{equation} \label{phi_ind}
\ket{\Psi_{d}^{N}}_{i}=\frac{1}{\sqrt{2d}}\sum_{j=1}^{d}(\ket{H_{j}}^{\otimes \frac{N}{2}}\ket{H_{j+1}}^{\otimes \frac{N}{2}}+\ket{V_{j}}^{\otimes \frac{N}{2}}\ket{V_{j+1}}^{\otimes \frac{N}{2}}),
\end{equation}
where notations  $\ket{H_{d+1}}:=\ket{H_1}$ and $\ket{V_{d+1}}:=\ket{V_1}$.
This system exhibits symmetry on the indexes $\{1,2,\ldots,d\}$ under certain cyclic permutations. 
Let us consider a cyclic permutation denoted by $\sigma$, which reorders elements in a sequence by shifting each element to the position of its successor, with the final element returning to the initial position. 
Mathematically, this operation is represented by $\sigma(j) = j + 1$ for $j \neq d$, and $\sigma(d) = 1$. This process preserves the sequential order of elements in a cyclical manner. 
Notably,  Eq.~\eqref{phi_ind} remains unchanged under the action of $\sigma$ and its integral powers.
As the photons propagate through the phases situated at the $d$ nodes, the resulting output state becomes
\begin{align} \label{phi_ond}
\ket{\Psi_{d}^{N}}_{o}&=\frac{1}{\sqrt{2d}}\sum_{j=1}^{d}(\ket{H_{j}}^{\otimes \frac{N}{2}}\ket{H_{j+1}}^{\otimes \frac{N}{2}}\nonumber \\
&+e^{i\frac{N}{2}(\phi_{j}+\phi_{j+1})}\ket{V_{j}}^{\otimes \frac{N}{2}}\ket{V_{j+1}}^{\otimes \frac{N}{2}}),
\end{align}
where $\phi_{d+1}:=\phi_1$. Note that the evolution of this system does not break its symmetry.

\subsection{The singularity of the QFIM}

Using Eq.~\eqref{fq_def}, we obtain the QFIM ${\cal F}_{0}^{(Q)}$ of the original set of variables, which elements are:
\begin{align}
    ({\cal F}_{0}^{(Q)})_{jj}&=\frac{N^{2}}{d^{2}}(d-1),\nonumber \\
    ({\cal F}_{0}^{(Q)})_{jj+1}&=\frac{N^{2}}{d^{2}}(\frac{d}{2}-1), \nonumber \\
    ({\cal F}_{0}^{(Q)})_{jj+k}&=-\frac{N^{2}}{d^{2}},\qquad(k\geq2).
\end{align}
The determinant of ${\cal F}_{0}^{(Q)}$ can be calculated using the Leibniz formula
\begin{equation}
    \det({\cal F}_{0}^{(Q)}) = \sum_{\sigma \in S_n} \text{sgn}(\sigma) \prod_{j=1}^{n} ({\cal F}_{0}^{(Q)})_{j, \sigma(j)}.
\end{equation}
Consider any row of ${\cal F}_{0}^{(Q)}$, which contains $ {\cal F}_{j1}^{(Q)}$, $ {\cal F}_{j2}^{(Q)}$, $\ldots$, ${\cal F}_{jd}^{(Q)}$. The combination $\sum_{k=1}^{d}(-1)^{k}({\cal F}_{0}^{(Q)})_{jk}=0$, indicating that the columns of ${\cal F}_{0}^{(Q)}$ form a linearly dependent set. Consequently, the determinant of ${\cal F}_{0}^{(Q)}$ is $0$, implying that ${\cal F}_{0}^{(Q)}$ is singular.

\subsection{Finding the irrelevant variable}

To examine the system's degree of freedom, we consider a linear combination of the variables defined as $\theta_0:=\frac{1}{d}\sum_{j=1}^{d}(-1)^j\phi_j:=\overrightarrow{c_0}\cdot\overrightarrow{\phi}$. 
We demonstrate the independence of the system's evolution on $\theta_0$ by considering the following variable transformation:
\begin{equation} \label{transformation}
(\theta_{0},\theta_{1},\theta_{2},\cdots,\theta_{d-1})^{T}=M_{c}\cdot(\phi_{1},\phi_{2},\phi_{3},\cdots,\phi_{d})^{T},
\end{equation}
where  $\theta_{1}=\overrightarrow{c_1}\cdot\overrightarrow{\phi}=\bar{\phi}$ being the average phase. Thus 
$\overrightarrow{c_1} = (\frac{1}{d},\cdots,\frac{1}{d})$. It's apparent that $\overrightarrow{c_1}\bot\overrightarrow{c_0}$. 
We then let $\theta_{i}=\overrightarrow{c_i}\cdot\overrightarrow{\phi}=\frac{1}{d}(\phi_{i-1}-\phi_{i+1})$ for $2 \leq i \leq (d-1)$. 
Note that $\overrightarrow{c_i}\bot\overrightarrow{c_0}$ and $\overrightarrow{c_i}\bot\overrightarrow{c_1}$ for $2 \leq i \leq (d-1)$. 
The transformation described can be defined by the transformation matrix 
\begin{equation} \label{mc}
    M_{c}=\left(\begin{array}{c}
\overrightarrow{c_{0}}\\
\overrightarrow{c_{1}}\\
\overrightarrow{c_{2}}\\
\overrightarrow{c_{3}}\\
\vdots\\
\overrightarrow{c_{d-2}}\\
\overrightarrow{c_{d-1}}
\end{array}\right)=\frac{1}{d}\left(\begin{array}{ccccccc}
-1 & 1 & -1 & 1 & -1 & \cdots & 1\\
1 & 1 & 1 & 1 & 1 & \cdots & 1\\
1 & 0 & -1 & 0 & 0 & \cdots & 0\\
0 & 1 & 0 & -1 & 0 & \cdots & 0\\
\vdots & \vdots & \vdots & \vdots & \vdots & \ddots & \vdots\\
0 & \cdots & 0 & 1 & 0 & -1 & 0\\
0 & \cdots & 0 & 0 & 1 & 0 & -1
\end{array}\right),
\end{equation}
where for $1 \leq i \leq d$, 
\begin{align*}
    (M_c)_{1i}&=\frac{1}{d}(-1)^i, \qquad
    (M_c)_{2i}=\frac{1}{d}, \\
    (M_c)_{ji}&=\frac{1}{d}[\delta(j-2,i)-\delta(j,i),\quad \text{for}\quad (3 \leq j \leq d)].
\end{align*}
Here the notation $\delta(x_1,x_2)$ refers to the kronecker delta, defined as $\delta(x_1,x_2)=1$ if $x_1=x_2$, and $\delta(x_1,x_2)=0$ otherwise.  
It's important to note that this transformation is not orthogonal. Specifically, not every pair of variables $\varphi_i$ and $\varphi_j$ within this set are orthogonal; for example, $\overrightarrow{c_{i}}\cdot\overrightarrow{c_{i+1}}=1/d^2$ for $2 \leq i \leq (d-2)$.
However, as long as this transformation is invertible, orthogonality is not a prerequisite, as our primary concern is for the quantum Fisher information matrix (QFIM) to be block diagonal. In the case of the transformation described in Equation~\eqref{mc}, the elements of the inverse transformation matrix $M_{c}^{-1}$ are
\begin{align}\label{mc_1}
    (M_c^{-1})_{i1}&=(-1)^i, \qquad
    (M_c^{-1})_{i2}=1, \nonumber \\
    (M_c^{-1})_{ij}&=d\delta[(-1)^i,(-1)^j]\{H(j-i)(1-\frac{j-2+\delta[(-1)^i,1]}{d}) \nonumber \\
    &-H(i-j)\frac{j-2+\delta[(-1)^i,1]}{d}\},\;\;\;\; \text{for}\;\;\; (3 \leq j \leq d).
\end{align}
In this context, $H(x)$ represents a modified Heaviside function, defined as $H(x)=1$ if $x\geq 0$, and $H(x)=0$ otherwise. 

We analyze the dependence of the system on the variable $\theta_0$
\begin{align} \label{independent}
\partial_{\theta_0}&\ket{\Psi_{d}^{N}}_{o}=\sum_{i=1}^{d}\frac{\partial {\phi_i}}{\partial {\theta_0}}\partial_{\phi_i}\ket{\Psi_{d}^{N}}_{o} \nonumber \\
&=i\frac{N}{2d}\sum_{i=1}^{d}(-1)^ie^{i\frac{N}{2}(\phi_{i-1}+\phi_{i})}\ket{V_{i-1}}^{\otimes \frac{N}{2}}\ket{V_{i}}^{\otimes \frac{N}{2}} \nonumber \\
&+i\frac{N}{2d}\sum_{i=1}^{d}(-1)^ie^{i\frac{N}{2}(\phi_{i}+\phi_{i+1})}\ket{V_{i}}^{\otimes \frac{N}{2}}\ket{V_{i+1}}^{\otimes \frac{N}{2}}.
\end{align}
Note that by substituting $i$ with $j-1$, the second term becomes the opposite of the first term, i.e.,
\begin{equation}\label{irr}
\partial_{\theta_0}\ket{\Psi_{d}^{N}}_{o}=0.
\end{equation}
This indicates that the system described in Eq.~\eqref{phi_ond} does not have a degree of freedom $d$ but at most $(d-1)$ instead. Notably, we prove that the degree of freedom is precisely $(d-1)$ by contradiction, showing that $\ket{\Psi_{d}^{N}}_{o}$ is dependent on any other linear combinations of $\{\phi_1,\ldots,\phi_d\}$.
Consider $\theta_x:=\overrightarrow{c_x}\cdot\overrightarrow{\phi}$, where $\cos(\overrightarrow{c_0},\overrightarrow{c_x})\neq\pm 1$, we have
\begin{align} \label{dependent}
\partial_{\theta_x}&\ket{\Psi_{d}^{N}}_{o}=\sum_{i=1}^{d}\frac{\partial {\phi_i}}{\partial {\theta_x}}\partial_{\phi_i}\ket{\Psi_{d}^{N}}_{o}\nonumber \\
&=i\frac{N}{2}\sum_{i=1}^{d}[(M_c^{-1})_{ix}+(M_c^{-1})_{i-1,x}]e^{i\frac{N}{2}(\phi_{i-1}+\phi_{i})}\ket{V_{i-1}}^{\otimes \frac{N}{2}}\ket{V_{i}}^{\otimes \frac{N}{2}}.
\end{align}
Note that when Eq.~\eqref{dependent} evaluates to zero, it implies that $(M_c^{-1})_{ix}=-(M_c^{-1})_{i-1,x}$ for all $i\in\{1,\ldots,d\}$, which indicates $\cos(\overrightarrow{c_0},\overrightarrow{c_x})=\pm 1$. 
This completes our proof regarding the system's degree of freedom, which is $(d-1)$.

\subsection{The Heisenberg QCRB for the average phase}

Therefore, we compute the quantum Fisher information matrix ${\cal F}^{(Q)}$ for the variables $\{\theta_1,\ldots,\theta_{d-1}\}$. This matrix has a rank of $(d-1)$, indicating its invertibility.
We show that such a transformation leads to a block diagonal QFIM of dimension $(d-1)$, where ${\cal F}^{(Q)}_{1i}=0$ for all $i \neq 1$. Consider any variables $\theta_x\,(x \neq 1)$, we have
\begin{align} \label{off_diagonal}
    {\cal F}^{(Q)}_{1x} &= - N^2\text{Re}\{\frac{1}{2d}\{\sum_{i=1}^{d}\frac{2}{d}[(M_c^{-1})_{ix}+(M_c^{-1})_{i-1,x}] \nonumber \\
    &- \frac{1}{4d^2} 2\sum_{i=1}^{d}[(M_c^{-1})_{ix}+(M_c^{-1})_{i-1,x}] \} \nonumber \\
    &= \frac{N^2}{d^2}\text{Re}\sum_{i=1}^{d}(M_c^{-1})_{ix}.
\end{align}
As given in Eq.~\eqref{mc_1},
\begin{equation}\label{mcsum}
    \sum_{i=1}^{d}(M_c^{-1})_{ix}=0,
\end{equation}
for all $2 \leq x \leq d-1$. Consequently, we establish the block diagonal structure of the quantum Fisher information matrix ${\cal F}^{(Q)}$.
Given that a block diagonal matrix can be inverted block by block, we can thereby derive the quantum Cramér-Rao bound for $\theta_1$ as
\begin{equation} \label{qcrb_nd}
    \Delta_q \theta_1\geq \sqrt{({\cal F}^{(Q)})^{-1}_{11}}=({\cal F}^{(Q)}_{11})^{-1/2}=N^{-1},
\end{equation}
which shows that the Heisenberg scaling holds for the quantum Cramér-Rao bound of the average phase, regardless of the number of the nodes $d$.

\section{Measurement and the Cramér-Rao bound}

Consider the same projective measurement on the $\sigma_x$ basis as in Eq.~\eqref{pjk_1}. The corresponding probability set now comprises $4d$ probabilities denoted as ${P_{N,jk}^{++},P_{N,jk}^{--},P_{N,jk}^{+-},P_{N,jk}^{-+}}$. 
For instance, $P_{N,d1}^{+-}$ represents the probability detecting the state $\ket{+_d}^{\otimes \frac{N}{2}}\ket{-_1}^{\otimes \frac{N}{2}}$.
The probabilities for the projective measurement are indicated as
\begin{align} \label{pjk_n}
P_{N,jk}^{++}=P_{N,jk}^{--}=\frac{1+\cos\frac{N}{2}(\phi_{j}+\phi_{k})}{4d},\nonumber \\
P_{N,jk}^{+-}=P_{N,jk}^{-+}=\frac{1-\cos\frac{N}{2}(\phi_{j}+\phi_{k})}{4d}.
\end{align}
Using the original basis, the elements in the FIM are calculated from
\begin{equation}\label{fmn_n}
    (F_0)_{mn}=2\sum_{j,k=1}^{d}(\frac{1}{P_{N,jk}^{\pm\pm}}\frac{dP_{N,jk}^{\pm\pm}}{d\phi_{m}}\frac{dP_{N,jk}^{\pm\pm}}{d\phi_{n}}+\frac{1}{P_{N,jk}^{\pm\mp}}\frac{dP_{N,jk}^{\pm\mp}}{d\phi_{m}}\frac{dP_{N,jk}^{\pm\mp}}{d\phi_{n}}),
\end{equation}
where
\begin{align}
    \frac{dP_{N,jk}^{\pm\pm}}{d\phi_{i}}&=-\frac{N}{2}\frac{\sin\frac{N}{2}(\phi_{j}+\phi_{k})}{4d}[\delta(i,j)+\delta(i,k)],\nonumber \\
    \frac{dP_{N<jk}^{\pm\mp}}{d\phi_{i}}&=\frac{N}{2}\frac{\sin\frac{N}{2}(\phi_{j}+\phi_{k})}{4d}[\delta(i,j)+\delta(i,k)].
\end{align}
Using Eq.~\eqref{fmn_n}, we obtain
\begin{equation}
    (F_0)_{mn}=\frac{N^{2}}{4d}[2\delta(m,n)+\delta(m-1,n)+\delta(m,n-1)].
\end{equation}
Since the equation $F_0x=0$ has nonzero solutions, such as
\begin{equation}
F_0\left(1,-1,1,-1,\ldots,1,-1\right)^{T}=0,
\end{equation}
it follows that the FIM $F_0$ is singular. 

To address the singularity, we apply the same transformation described in Eq.~\eqref{transformation}. The new FIM $F$ is then expressed as
\begin{equation}\label{fmn_n2}
    (F)_{mn}=2\sum_{j,k=1}^{d-1}(\frac{1}{P_{N,jk}^{\pm\pm}}\frac{dP_{N,jk}^{\pm\pm}}{d\theta_{m}}\frac{dP_{N,jk}^{\pm\pm}}{d\theta_{n}}+\frac{1}{P_{N,jk}^{\pm\mp}}\frac{dP_{N,jk}^{\pm\mp}}{d\theta_{m}}\frac{dP_{N,jk}^{\pm\mp}}{d\theta_{n}}).
\end{equation}
We then demonstrate that $F$ is block-diagonal. In the sense that $F_{1i}=F_{i1}=0$ for all $i\neq 1$. Firstly, we have
\begin{align}
    \frac{dP_{N,n,n+1}^{\pm\pm}}{d\theta_{i}}&=\sum\frac{dP_{N,n,n+1}^{\pm\pm}}{d\phi_{k}}\frac{d\phi_{k}}{d\theta_{i}} \nonumber \\
    &=\frac{dP_{N,n,n+1}^{\pm\pm}}{d\phi_{n}}\frac{d\phi_{n}}{d\theta_{i}}+\frac{dP_{N,n,n+1}^{\pm\pm}}{d\phi_{n+1}}\frac{d\phi_{n+1}}{d\theta_{i}} \nonumber \\
    &=-\frac{N}{2}\frac{\sin\frac{N}{2}(\phi_{n}+\phi_{n+1})}{4d}(\frac{d\phi_{n}}{d\theta_{i}}+\frac{d\phi_{n+1}}{d\theta_{i}}).
\end{align}
Similarly
\begin{equation}
    \frac{dP_{N,n,n+1}^{\pm\mp}}{d\theta_{i}}=\frac{N}{2}\frac{\sin\frac{N}{2}(\phi_{n}+\phi_{n+1})}{4d}(\frac{d\phi_{n}}{d\theta_{i}}+\frac{d\phi_{n+1}}{d\theta_{i}})
\end{equation}
Using Eq.~\eqref{mcsum}, we obtain
\begin{align}
    F_{1i}&=F_{i1}=\frac{N^{2}}{4d}\sum_{n}(\frac{d\phi_{n}}{d\theta_{1}}+\frac{d\phi_{n+1}}{d\theta_{1}})(\frac{d\phi_{n}}{d\theta_{i}}+\frac{d\phi_{n+1}}{d\theta_{i}}) \nonumber \\
    &=\frac{N^{2}}{2d}\sum_{n}(\frac{d\phi_{n}}{d\theta_{i}}+\frac{d\phi_{n+1}}{d\theta_{i}})=\frac{N^{2}}{d}\sum_{n}(M_{c}^{-1})_{ni}=0,
\end{align}
and
\begin{equation}\label{classical_11}
    F_{11}=\frac{N^{2}}{d}\sum_{n}(M_{c}^{-1})_{n1}=N^{2}.
\end{equation}
Thus, by performing projective measurements, we can experimentally attain the Quantum Cramér-Rao Bound for estimating the average phase, where this straightforward process is as efficient and accurate as theoretically possible.

\section{The relation between measured Weak Cramér-Rao bound and exact Cramér-Rao bound}

We have shown in previous sections that for the multiphoton states with projective measurements, the weak Cramér-Rao bound and exact bound coincide.
Now, we demonstrate that for any nonsingular, Hermitian, and positive definite matrix $S$, the following holds:
\begin{equation}\label{a1}
   \frac{(\alpha^{T}\alpha)^{2}}{\alpha^{T}S\alpha}\leq\alpha^{T}S^{-1}\alpha .
\end{equation}
By applying the Cauchy-Schwarz inequality, we have
\begin{equation}
   (\beta^\dagger\gamma)^{2}\leq(\beta^\dagger\beta)(\gamma^\dagger\gamma),
\end{equation}
for any vectors $\beta$ and $\gamma$. Given that $S$ is positive definite, we set $S^{1/2}\alpha\coloneqq\beta$, and $S^{-1/2}\alpha\coloneqq\gamma$. Substituting these into the inequality, we obtain
\begin{equation}   
[\alpha^{T}(S^{-1/2})^\dagger S^{1/2}\alpha]^{2}\leq[\alpha^{T}(S^{1/2})^\dagger S^{1/2}\alpha][\alpha^{T}(S^{-1/2})^\dagger S^{-1/2}\alpha] .
\end{equation}
Since $S$ is Hermitian, $S^{-1/2}$ is also Hermitian. Thus, we simplify the expression to
\begin{equation}   
(\alpha^{T}\alpha)^{2}\leq(\alpha^{T}S\alpha)(\alpha^{T}S^{-1}\alpha) .
\end{equation}
As $S$ is positive definite, multiplying both sides by $(\alpha^{T}S\alpha)^{-1}$ yields Eq.~\eqref{a1}. 
The condition for equality is satisfied when $\beta$ and $\gamma$ are proportional, which corresponds to $\alpha$ and $S\alpha$ being proportional. As a special case of Eq.~\eqref{a1}, we have
\begin{equation}
(S_{11})^{-1}\leq(S^{-1})_{11}.
\end{equation}
\noindent This is the relation between the weak and exact Cramér-Rao bounds for the average phase, assuming that the $11$ element of the nonsingular Fisher information matrix $[{\cal F}^{(Q)}, F]$ represents the meaurement of average phase.

\section{Discussions}

In this work, we addressed a critical obstacle in distributed sensing using multiphoton: the singularity issue within the Fisher information matrix arising from the interdependence among phase variables when employing polarization Bell states. 
To confront this challenge, we proposed innovative variable transformations, a pivotal step towards obtaining the quantum Cramér-Rao bounds (QCRB). 
Going beyond the scope of prior studies, we expanded our analysis from the foundational scenario outlined by Kim et al. (where N=2 and d=4) to arbitrary even numbers N and d. 
We demonstrated this system's ability to achieve Heisenberg scaling on estimating the average phase. 
This validation underscores the effectiveness of distributed sensing techniques, emphasizing their applicability for analyzing linear combinations of distributed variables. 
Understanding the interplay between these combinations and their independent constituents provides the potential for more advanced and efficient sensing methodologies.

\section*{ACKNOWLEDGMENTS}
The authors thank Dr. Luiz Davidovich and Dr. Shouda Wang for the invaluable discussions. The authors are grateful for the support of Air Force Office of Scientific Research (Award No. FA-9550-20-1-0366), the Robert A. Welch Foundation (A-1943-20240404), and the DOE Award (DE-AC36-08GO28308).

\bibliographystyle{ieeetr}

\end{document}